\documentclass[compsoc, conference, a4paper, 10pt, times]{IEEEtran}

\usepackage{cite}
\usepackage{amsmath,amssymb,amsfonts}
\usepackage{algorithmic}
\usepackage{graphicx}
\usepackage{textcomp}
\usepackage{xcolor}
\usepackage{booktabs}
\usepackage[hidelinks]{hyperref}
\usepackage{listings}
\usepackage[T1]{fontenc}
\usepackage{todonotes}

\begin{document}

\title{Towards a better labeling process for network security datasets}


\author{\IEEEauthorblockN{Sebastian Garcia}
\IEEEauthorblockA{\textit{Department of Computer Science} \\ \textit{Czech Technical University} \\
\textit{Prague, Czech Republic}\\
sebastian.garcia@agents.fel.cvut.cz \\}
\and
\IEEEauthorblockN{Veronica Valeros}
\IEEEauthorblockA{\textit{Department of Computer Science}  \\ \textit{Czech Technical University} \\
\textit{Prague, Czech Republic}\\
veronica.valeros@fel.cvut.cz}
}

\maketitle

\begin{abstract}
Most network security datasets do not have comprehensive label assignment criteria, hindering the evaluation of the datasets, the training of models, the results obtained, the comparison with other methods, and the evaluation in real-life scenarios. There is no labeling ontology nor tools to help assign the labels, resulting in most analyzed datasets assigning labels in files or directory names.
This paper addresses the problem of having a better labeling process by (i) reviewing the needs of stakeholders of the datasets, from creators to model users, (ii) presenting a new ontology of label assignment, (iii) presenting a new tool for assigning structured labels for Zeek network flows based on the ontology, and (iv) studying the differences between generating labels and consuming labels in real-life scenarios. We conclude that a process for structured label assignment is paramount for advancing research in network security and that the new ontology-based label assignation rules should be published as an artifact of every dataset.
\end{abstract}

\section{Introduction}
Assigning labels correctly to a network security dataset is paramount for the development of detection methods~\cite{guerra_datasets_2022}. It is known that without labels, and without \textit{good quality} labels, supervised machine learning methods can not be trained, and unsupervised methods can not be verified. Without a known structured process for assigning labels, datasets and models can not be compared or improved. However, the lack of a good-quality \textit{labeling process} also decreases the usefulness of a dataset, and this problem is not deeply studied.

In many domains, and network security is not an exception, it is very hard to assign ground-truth labels, with studies pointing out that having labels is not enough for a good dataset~\cite{guerra_datasets_2022}. It is a hard process not only because of the needed confidence and skills of the expert assigning the labels but also due to the lack of a process to design and implement labels. Labeling requires both deep comprehension of the process that generated the traffic, a good labeling process, and a good tool to put the labels. Correct assignment of labels can take as much time as the capture of the traffic itself. 

It is not surprising then that some datasets lack: good descriptive labels, only include binary labels, lack a deep explanation of how the labels were assigned, or lack labels at all. Most commonly, datasets assign the same label to different concepts resulting in the impossibility to compare them. This common unstructured approach that the community has taken forces machine learning model creators to error-prone and intensive processing to use each dataset.

Previous work in the creation of ontologies for cybersecurity mainly focused on high-level problems, such as threat hunting, incidence response, or data exchange~\cite{bromander_semantic_nodate}. No ontology was identified focusing on the labeling process of network security datasets, although some ontologies include a network perspective~\cite{syed_uco_nodate}. Only two previous tools were found to label network flows, and none used an ontology.

This paper explores the problem of a process for labeling network datasets in four parts: the need and importance of labels; the criteria to assign labels using an ontology; a tool to assign them; and how they are consumed in real-life scenarios. To facilitate the discussion, a simple use case is shown across the paper: how to label a port scan?

In the first part,~\autoref{sec-need-labels}, we analyze the need for labels, including who needs them, for what purpose, and which are the specific characteristics of those needs. We identify the stakeholders of labels' consumption, the cost of assigning labels, and the technical difficulties underlying the process. We explore the usefulness of binary labels compared with multi-category labels. 

In the second part,~\autoref{sec-ontology} we argue the need for an ontology to assign structured labels and we propose a new ontology. The ontology is loosely based on parts of the MITRE ATT\&CK framework~\cite{blake_e_strom_mitre_2018}. 

In the third part,~\autoref{sec-tool} we present a Python tool to assign labels to network security flows. It reads a configuration file and automatically assigns labels to \textbf{all} the files generated by the Zeek network security monitoring tool~\cite{noauthor_zeek_nodate} (not only the conn.log file). The configuration file follows the ontology and uses a simple rule-based system.

In the fourth part,~\autoref{sec-using-labels}, we explore the uses of labels by the stakeholder of datasets, that needs to compare and evaluate third-party datasets and models in real-life scenarios. This requires knowledge of when the attack happened, when the detection happened, and which components of the traffic were involved in the detection. The analysis highlights the difference between producing labels and consuming labels.


We conclude that the proposed analysis, ontology, and tool can help the security community greatly improve the production of datasets and the comparison and evaluation of models. In particular, we conclude that the configuration file used for label assignments should be a \textbf{published artifact} of datasets to allow reproducibility and comparison.


\section{Previous work}
\label{sec-prevwork}

Even though previous work on network datasets often explored the concept of labels in datasets~\cite{ring_survey_2019}, the exploration was mostly done at a high level checking for example if any label was present and not focusing on the properties of labels. Other domains, such as computer vision, have worked extensively on this problem and are already aware of the need for correct, organized, and confident labels~\cite{northcutt_confident_2021}.

\subsection{Ontologies for label representation}
Extensive work has been done in semantic analysis of the cybersecurity domain, including ontologies for threat intelligence and attack graphs~\cite{bromander_semantic_nodate}\cite{aviad_semantic_2015}\cite{mavroeidis_cyber_2017}\cite{zhang_review_2020}\cite{valja_automating_2020}\cite{kenaza_toward_2016}\cite{undercoffer_modeling_2003}. However, none of these ontologies is focused on the procedure to assign labels for network traffic or in the work on datasets. Most ontologies focus on the larger picture of threat hunting and security incident cataloging but only briefly discuss labels. Network traffic in security datasets has not been the focus of previous ontologies.

An example of a good high-level cybersecurity ontology is UCO~\cite{syed_uco_nodate}. UCO is focused on cyber situational awareness and data sharing, putting together many aspects of security. However, there is no specific part for labels in the datasets since it is out of scope. 

\subsection{Network security datasets and labels}
Prior work has highlighted the difficulty of labeling network traffic and the lack of reproducibility of the labeling methodology~\cite{guerra_datasets_2022}. In addition, published datasets sometimes contain errors that require significant effort to correct~\cite{liu_error_2022}. For this work, we analyzed existing network security datasets in terms of \textit{structure} and \textit{location} of labels. This analysis shed light on the lack of awareness of the problem and allowed us to perform a simple comparison of labeling approaches. The label location assignment criterion in the current datasets can be divided into the following styles:
\begin{itemize}
    \item A-In packets.
    \item B-In flows samples.
    \item C-In file names.
    \item D-In directory names.
    \item E-In metadata.
    \item F-No labels.
\end{itemize}

Style \textit{A-In packets} means that labels were assigned inside the packets. \textit{B-In flow samples} means that labels were assigned flow by flow. \textit{C-In file names} means that labels are assigned in the dataset file names. \textit{D-in directory names} means that labels were assigned to the names of directories and, by inference, to all the files inside of any type. \textit{E-In metadata} means that labels were only assigned in readme files, webpages, documentation, papers, etc. \textit{F-No labels} means that despite mentions of assigned labels, they can not be found or used for various reasons. In the case of multiple styles, only the more precise method is reported, C.

To verify our work we downloaded, described, and analyzed several datasets in network security. Our list of 36 datasets is not exhaustive, but as far as we know is the longest list of network security datasets to date.

\subsubsection{SUEE 2017}
This dataset contains being and attack traffic~\cite{thomas_lukaseder_2017-suee-data-set_2022}. The labeling style is \textit{E-In metadata}.

\subsubsection{CSE-CIC-IDS2018}

This dataset contains network and host data~\cite{sharafaldin_toward_2018}. The labeling style is \textit{B-In Flows samples}.

\subsubsection{CIC IoT 2022}
This dataset contains benign IoT devices and some attacks~\cite{dadkhah_towards_2022}. Unfortunately, it is not clear how the labels were assigned or where they were assigned. The labeling style is \textit{F-No labels}.

\subsubsection{CIC IoT Enriched}
This dataset is a combination of two IoT datasets, Bot-IoT and Ton-IoT, that were extended with more features~\cite{erfani_feature_2021}. This enriched dataset did not describe, name, add or modify any original label. The labeling style is \textit{F-No labels}.

\subsubsection{ISCX-IDS-2012}
This dataset contains anomaly-focused synthetic data~\cite{shiravi_toward_2012}. The labeling style is \textit{B-In flows samples}.

\subsubsection{CIC-AAGM2017}
This dataset contains Android benign, adware, and malware applications and traffic~\cite{lashkari_towards_2017}. The labeling style is \textit{D-In directory names}.

\subsubsection{CIC-MalAnal2017}
This dataset contains Android benign and malware applications and traffic~\cite{lashkari_toward_2018}. The labeling style is \textit{B-In flow samples.}

\subsubsection{CIC-InvesAndMal2019}
This dataset contains benign and malicious Android applications and traffic~\cite{taheri_extensible_2019}. The labeling style is \textit{B-In flows samples}. 

\subsubsection{CIC-Bell-DNS-EXF-2021}
This dataset contains DNS exfiltration attacks and benign DNS requests~\cite{mahdavifar_lightweight_2021}. The labeling style is \textit{D-In directory names}.

\subsubsection{CIRA-CIC-DoHBrw-2020}
This dataset contains benign and malicious DoH traffic~\cite{montazerishatoori_detection_2020}. The labeling style is \textit{C-In file names}.

\subsubsection{CIC-DDoS2019}
This dataset contains synthetic benign and real DDoS attacks~\cite{sharafaldin_developing_2019}. The labeling style is \textit{B-In flow samples}.

\subsubsection{CIC-IDS 2017}
This dataset contains synthetic benign traffic~\cite{sharafaldin_toward_2018}. The labeling style is \textit{B-In flow samples}.

\subsubsection{CIC DoS dataset}
This dataset mixed DDoS attacks with the benign part of the ISCX-IDS dataset~\cite{jazi_detecting_2017}. The label assignment style is \textit{E-In metadata}.

\subsubsection{ISCX VPN 2016}
This dataset contains VPN and non-VPN traffic~\cite{draper-gil_characterization_2016}. The labeling style is \textit{F-No labels}.

\subsubsection{ISCX Tor 2016}
This dataset contains TOR and non-TOR traffic~\cite{habibi_lashkari_characterization_2017}. The labeling style is \textit{D-In directory names}.

\subsubsection{ISOT Botnet}
This dataset is a combination of other malicious datasets~\cite{saad_detecting_2011}. The labeling style is \textit{D-In metadata}.

\subsubsection{ISOT Ransomware Detection}
This dataset contains ransomware and benign traffic~\cite{isot_research_lab_botnet_nodate}. The label assignment style is \textit{D-In directory names}.

\subsubsection{ISOT HTTP Botnet}
This dataset is a mix of other datasets and original captures~\cite{alenazi_holistic_2017}. The assignment style is \textit{D-In directory names}.

\subsubsection{ISCX Bot 2014}
This dataset is a mix of datasets~\cite{biglar_beigi_towards_2014}. The labeling style is \textit{E-In metadata}.

\subsubsection{TON\_IoT}
This dataset is a mix of datasets for host and network traffic~\cite{booij_ton_iot_2022}. The labeling style is \textit{B-In flow samples}.

\subsubsection{Bot-IoT}
This dataset contains normal and malware traffic~\cite{koroniotis_towards_2018}. The labeling style is \textit{B-In flow samples}.

\subsubsection{IoTBot-IDS}
This dataset contains normal and botnet traffic~\cite{ashraf_iotbot-ids_2021}. The labeling style is \textit{D-In directory names}.

\subsubsection{UNSW-BNB15}
This dataset contains synthetic attacks and synthetic normal traffic~\cite{moustafa_unsw-nb15_2015}. The labeling style is \textit{B-In flows samples}.

\subsubsection{NSL-KDD}
This dataset contains an improvement of normal and anomaly records for the KDD99 dataset~\cite{tavallaee_detailed_2009}. Unfortunately, the labeled TXT and ARFF files do not contain flows. The labeling strategy is \textit{F-No labels}.

\subsubsection{CTU-13}
This dataset contains malware, benign, and background traffic~\cite{garcia_empirical_2014}. The labeling style is \textit{B-In flows}.


\subsubsection{CTU IOT-23}
This dataset contains malware and benign traffic~\cite{garcia_iot-23_2020}. Labels were assigned using the Flaber tool~\cite{agustin_parmisano_flaber_2022} to flows, so the style is \textit{B-In flows.}

\subsubsection{CTU HORNET-40}
This dataset contains malicious activity. It includes the sub-datasets CTU-Hornet-7 (the first 7 seven days), and CTU-Hornet-15 (the first 15 days). The labeling style is \textit{C-In file names}.

\subsubsection{CTU Android Mischief}
This dataset contains malicious and benign traffic of RAT Android applications~\cite{babayeva_android_2021}. The labeling style is \textit{B-In file names}.

\subsubsection{CTU DNS Threats}
This dataset contains DGA and benign DNS domain names~\cite{palau_detecting_nodate}. The labeling style is \textit{B-In flow samples}.

\subsubsection{CTU AIP Attacks 2022}
This dataset contains attacks to honeypots~\cite{bogado_ctu-aip-attacks-2022_2023}. The labeling style is \textit{C-In file names}.

\subsubsection{CTU Malware Capture Facility Project}
The malware capture facility project is a collection of many malware, benign, mixed, and attack captures~\cite{stratosphere_laboratory_malware_nodate}. The list includes 410 individual malware infections, 2 manual attacks, 10 mixed simultaneous malware and benign, 26 verified benign traffic, 4 real IoT honeypots, 20 IoT malware infections, and 5 IRC botnets. The labeling style is \textit{D-In directory names}.

\subsubsection{CUPID}
This dataset contains attack traffic~\cite{lawrence_cupid_2022}. Unfortunately, it is not available for download but is alleged to be of labeling style \textit{E-In metadata}.

\subsubsection{DAPT 2020}
This dataset contains simulated APT attacks
~\cite{myneni_dapt_2020}. The labeling style is \textit{E-In metadata}.

\subsubsection{IoT Sentinel}
This dataset contains benign IoT traffic~\cite{miettinen_iot_2017}. The labeling style is \textit{D-In directory names}.

\subsubsection{LANL ARCS Dataset}
This dataset~\cite{turcotte_unified_2018} contains benign host and network traffic. The labeling style is \textit{E-In Metadata}.

\subsubsection{N-BaIoT}
This dataset contains botnet attack traffic~\cite{meidan_n-baiot_2018}. The labeling style is \textit{C-In file names}.


\begin{table}[tb]
\tiny
\renewcommand{\arraystretch}{1.5}
\caption{Overview of network security datasets labeling styles. A-In packets, B-In flows samples, C-In file names, D-In directory names, E-In metadata, and F-No labels.}
\begin{center}
\begin{tabular}{l r|r|r|r|r|r}
\toprule
\textbf{Dataset} &  \textbf{A} &  \textbf{B} &  \textbf{C} &  \textbf{D} &  \textbf{E} & \textit{F} \\
\midrule
SUEE 2017                       &   &   &   &   & \checkmark &  \\
CSE-CIC-IDS2018                 &   & \checkmark  &   &   &   &  \\
CIC IoT 2022                    &   &   &   &   &   & \checkmark \\
CIC IoT Enriched                &   &   &   &   &   &  \checkmark  \\
ISCX-IDS-2012                   &   &  \checkmark &   &   &   &  \\
CIC-AAGM2017                    &   &   &   & \checkmark  &   &  \\
CIC-MalAnal2017                 &   & \checkmark  &   &   &   &  \\
CIC-InvesAndMal2019             &   & \checkmark  &   &   &   &  \\
CIC-Bell-DNS-EXF-2021           &   &   &   & \checkmark  &   &  \\
CIRA-CIC-DoHBrw-2020            &   &   & \checkmark   &   &   &  \\
CIC-DDoS2019                    &   & \checkmark  &   &   &   &  \\
CIC-IDS2017                     &   & \checkmark  &   &   &   &  \\
CIC DoS dataset                 &   &   &   &   &  \checkmark &  \\
ISCX VPN 2016                   &   &   &   &   &   & \checkmark \\
ISCX Tor 2016                   &   &   &   & \checkmark  &   &  \\
ISOT Botnet                     &   &   &   & \checkmark  &   &  \\
ISOT Ransomware Detection       &   &   &   & \checkmark  &   &  \\
ISOT HTTP Botnet                &   &   &   & \checkmark  &   &  \\
ISCX Bot 2014                   &   &   &   &   & \checkmark  &  \\
TON\_IoT                        &   & \checkmark  &   &   &   &  \\
BOT-IOT                         &   & \checkmark  &   &   &   &  \\
IoTBot-IDS                      &   &  &  & \checkmark &   &  \\
UNSW-BNB15                      &   & \checkmark  &   &   &   &  \\
NSL-KDD                         &   &   &  &   &   &  \checkmark \\
CTU-13                          &   & \checkmark &   &   &   &  \\
CTU IOT-23                      &   & \checkmark  &   &   &   &  \\
CTU HORNET-40                   &   &   &  \checkmark  &   &   &  \\
CTU Android Mischief            &   &   &  \checkmark &   &   &  \\
CTU DNS Threats                 &   &   &   & \checkmark  &   &  \\
CTU AIP Attacks 2022            &   &   &  \checkmark &   &   &  \\
CTU Malware Capture Facility Project    &   &   &   &  \checkmark  &   &  \\
DAPT 2020                       &   &   &   &   &  \checkmark &  \\
CUPID                           &   &   &   &   &  \checkmark &  \\
IoT Sentinel                    &   &   &   &   &  \checkmark &  \\
LANL ARCS                       &   &   &   &   & \checkmark  &  \\
N-BaIoT                         &   &   & \checkmark  &   &   &  \\
\bottomrule     
\end{tabular}
\label{tab:datasets}
\end{center}
\end{table}

\section{The needs for network security labels}
\label{sec-need-labels}

Our analysis of the process for assigning labels starts with a discussion of why we need labels, which kind, and for what purpose. The most common need for labels seems to be to detect malicious traffic with machine learning models, and it is a problem of \textbf{detection} and not \textbf{classification} since most researchers seem to mostly want to distinguish \textit{malicious} from \textit{benign}\footnote{We are not differentiating between \textit{benign} and \textit{normal}, or \textit{malicious} and \textit{malware} labels since the paper is not about how to correctly choose labels.} traffic. This is because when the model is deployed in a real network, the defender will want to block (or not) the malicious traffic. The decision tends to be binary; therefore, the labels are too. Binary labels are the most common type of label because they are cheaper to \textit{decide}, need less expertise to assign, and are therefore less error-prone.

The need for more detailed labels only appears when there is a need to detect very specific behaviors. In these cases, the label \textit{malicious} is often not enough~\cite{engelen_troubleshooting_2021}, and more detailed labels are required, such as \textit{botnet}, \textit{spam}, \textit{ddos}, \textit{exfiltration}, or even \textit{mirai version 5}. As shown in~\autoref{tab:datasets}, not all datasets have detailed labels. From the 36 datasets reviewed, 11 had labels in \textit{B-In flows samples}, 9 in \textit{D-In directory names}, 7 in \textit{E-In metadata}, 5 in \textit{C-In file names}, and 4 were found as \textit{F-No labels}. No dataset was found to have labels in \textit{A-In packets}.


More importantly, the need for labels is strongly tied to the training, evaluation/testing, performance improvement, and explanation of machine learning methods. 

The need for labels in training is understood; without labels, no supervised method can be created. For unsupervised methods, it is also known that there are techniques that use labels to help clustering obtain better results~\cite{ghosh_supervised_2020}.

The need for labels for evaluation/testing is also usually understood; labels are needed to compute metrics on the output of the model and, in the case of unsupervised methods, to verify the coherence of the results.

The need for labels for performance improvement is less clear. The performance may not only be the accuracy but the detection of a specific type of attack or malware. Less commonly, it can also be the reduction of false positives by focusing on different classes of benign traffic. Here, labels are critical to transitioning from a first-stage binary detection model into a complex detection system in production networks.

Finally, the field of XAI (explainable AI) has a great need for labels. From techniques to explaining which features are more important by methods such as SHAP~\cite{roshan_utilizing_2021}, to the use of adversarial labeled counterfactual examples~\cite{kuppa_adversarial_2021}.

\subsection{Labels' stakeholders}
To fully identify the need for labels, it is vital to clearly identify the stakeholders of the labeling process and consumption. We identify three main stakeholders, (i) the dataset creator, (ii) the model creator, and (iii) the model consumer. The same person can be many simultaneous stakeholders.

The dataset creator is in charge of designing the labels, isolation the traffic, and assigning labels to the dataset so they can be consumed. They need expertise in the field of attacking and defense, traffic analysis, and also to be familiar with the many variants of benign traffic. Creators incur most of the cost for label assignment. 

The model creator is identified as the machine learning engineer or data scientist that designs, trains, and evaluates the models. Model creators need labels and frequent contact with datasets creators to iterate about the different needs. This is related to the \textit{continuous training} phase of an MLOps pipeline~\cite{kreuzberger_machine_2022}.

The model consumer includes the use of the model in a real scenario (production) and the use of the model for comparison or evaluation purposes. The model consumer needs labels to compare the model with other models, to verify if the model will be useful in real scenarios, and how its performance changes. This last use case is described in~\autoref{sec-using-labels}, and it implies that the evaluations of the model creators are not the same as the evaluations in real-life scenarios.

The identification of stakeholders and types of labels with different levels of detail raises the question of \textit{what is being labeled}? 

\subsection{Levels of labels' assignments}
We identify four levels to which labels can be assigned: (i) packets, (ii) flows of any kind, (iii) IP addresses, and (iv) users. Assigning labels to packets is the most precise method. This is because a malware-infected computer may generate packets \textbf{not created} by the malware process but by the operating system. Examples of such packets are ARP, multicast, and broadcast. Therefore, if a computer were infected with malware, \textit{not} all its packets should be labeled \textit{malicious}. If the same label is used for the whole PCAP file, then machine learning models may learn to detect operating system benign traffic as malicious.

The main disadvantage of labeling packets is \textit{where} to put the labels since no PCAP format officially supports them. The best solution so far is to add them in the comment sections of PCAPNG packets~\cite{gomaa_pcapng-parser_2023}.

In the case of network flows, labels can be very precise, allowing to distinguish malicious flows from benign flows in the same capture. All the packets in one flow are usually related to the same behavior. A problem with this approach is that to label flows, it is necessary to have flow files generated from the raw packet captures, extending the size of the dataset and including more files. Another problem is that there are many network flow standards, and it is impractical to generate them all (this is why most datasets always publish PCAP files if they can). Moreover, the same standard can generate different files depending on how it was configured. Despite these problems, labeling flows can be the most convenient way to publish labels since it has enough information for most machine learning detectors.

The need to label IP addresses comes for the use case of blocking the IP address of a computer that was detected as infected with malware. This is the case of many datasets labeling the IP addresses in a \textit{README} file. However, only labeling IP addresses is probably too generic and does not allow for much granularity during the training of specific models.

The last case of labeling users comes from the use case of blocking users in organizations where the same user has many IP addresses and devices, using them simultaneously. In an example scenario of a user having three devices: a laptop, a phone, and a tablet, and all of them have IPv4 and IPv6 addresses, a single user may have up to 6 IP addresses.

\subsection{What is labeled?}
When working at a packet or flow level, the label applies to the whole unit, for example, the whole flow is labeled as \textit{benign}. However, this is misleading because we are not saying which part of the flow is benign and why. Is the flow benign because the source IP is benign? Or because the destination IP is white-listed? Is the label related to the content? This idea of labels \textit{per feature} was already explored in other domains~\cite{druck_active_2009}.

For example, knowing that ICMP can be used for many purposes if an ICMP flow is labeled as \textit{port scan}, is it because it was sent by malware searching computers with ICMP? or because it is the answer from a benign IP address using ICMP as an answer to a closed UDP port? 

No dataset that we know of, includes a clear definition of what is the concept that is being labeled. This need is addressed in~\autoref{sec-ontology} where an ontology is defined to give researchers the opportunity to label network data with higher precision.

\subsection{The cost of labels}
One of the reasons why most labels are binary and why labels appear in file names is the high cost of labeling data. This is a well-known problem of machine learning in general. In network security, the cost involves (i) controlling the experiment to only generate the traffic that is desired, (ii) having control of the technology to capture what is desired, (iii) having the technical expertise to perform the desired action, (iv) having access to the malware files to execute, (v) having access to the desired benign traffic or benign behaviors, and (vi) having the expertise to analyze the traffic and find errors and biases. 

The cost of labeling a dataset has not been computed, but the estimated time needed is at least the same as that required to generate the dataset. This means that labeling duplicates the cost of the dataset generation. There are techniques to automatically label datasets using active learning~\cite{gao_consistency-based_2020}. However, these techniques have a strong bias and error, since they do not generate ground truth labels, just better labels prediction.

\subsection{Benign labels}
It is usually accepted that obtaining malicious traffic is hard. However, it is even more difficult to obtain benign traffic. This is because (i) it is protected by privacy regulations and can not be published, and (ii) because there is a lot of variability in benign behavior due to different operating systems, applications, types of users, bots, servers, time of day, etc. 

An inherent problem of benign traffic is that it is much harder to assign detailed labels than in malicious traffic. For example, should the label be \textit{benign} because it was not an attack? because the operating system generated it? because a web browser generated it? or because the website visited is well-known in the community? 

In addition, even though the device may be benign, it sometimes generates \textit{grey area} traffic, such as the one created by advertising, or by PUA (Probable Unwanted Applications) software.

\subsection{Use case: The need for labels in a port scan}
Through this paper, we reference the use case of a port scan to highlight the difficulties to label it. The use case refers to an example \textit{Nmap} (-sS) scan without host discovery (-Pn) and no service discovery (no -sV) to scan 10 TCP ports that are actually closed.

Using the example of a port scan, it would be easy to put a unique label \textit{port scan} to a PCAP file with traffic generated by \textit{Nmap}. However, \textit{Nmap} sends many packets that are not technically part of a port scan, such as checking if the IP address is active and working with ICMP or identifying the version of services using the respective protocols. This shows how complicated it can be to label a port scan and the dataset creator is required to know exactly which packets or flows were sent.

In most cases, detailed labels are needed as detection models need to train with specific concepts. Some examples of how many decisions need to be taken to correctly label a port scan are: 
\begin{enumerate}
    \item If a TCP port is closed during a port scan, the port-scanning tool may send only one SYN packet, but the operating system may \textit{resend} 2 more SYN packets. Are those 2 extra SYN packets part of the port scan? or part of the benign traffic generated by the operating system? 
    \item If a UDP port is closed, the operating system will send back an ICMP port unreachable packet. Is that ICMP part of the port scan?
\end{enumerate}

\section{An ontology for network security labels}
\label{sec-ontology}
The previous analysis of the need for better labeling helps us understand from which perspective we can label a dataset and how we can have labels that are generic enough to be used in binary classification but also detailed enough for most detection purposes. We, therefore, propose a new focused ontology for dataset creators to assign labels to their network traffic flows.

Section~\ref{sec-prevwork} describes previous cybersecurity ontologies created for particular purposes, such as attack attribution, threat modeling, and threat intelligence. These ontologies are too broad and complex for the narrow task of labeling a network security dataset in a way that is useful and usable. However, we incorporated some of their ideas in our ontology so it can be extended to be compliant with other ontology frameworks.

Our proposed ontology is designed to label network \textit{flows} (opposite to packets or IP addresses). The decision comes from the realization that flows are precise enough to capture the entire behavior of a connection and versatile enough to allow new columns in their text representation. Flows are also a very common representation of traffic in organizations because they are standardized, easy to understand and generate, and economical to store. Note that even though they are standardized, there are many standards, leading to compatibility issues. Our proposed ontology uses concepts from both the perspective of the attacker (e.g., which technique was used) and the perspective of the defender (e.g., state of the flow).

\begin{figure}[h]
    \caption{Proposed ontology to assign labels for network security flows}
    \centering
    \includegraphics[width=0.5\textwidth]{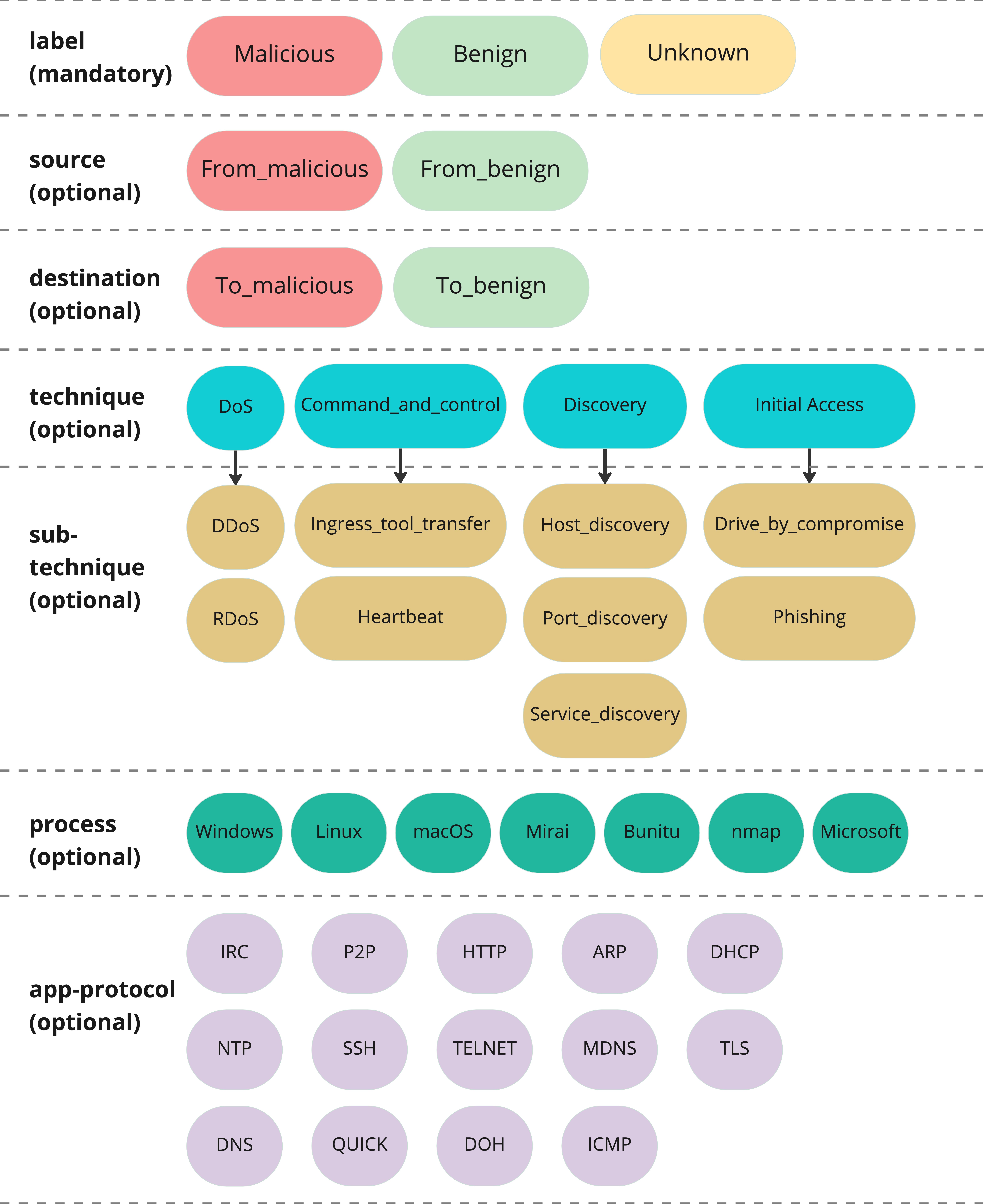}
    \label{fig-ontology}
\end{figure}

The ontology is presented as a structure in~\autoref{fig-ontology}. There are seven levels, the first is the mandatory main label of the flow, and the others are optional extra labels to be used as detailed labels. 

The first level is called \textit{label}, it is mandatory, and it is the label of the whole flow. This level contains three items:
\begin{itemize}
    \item \textbf{Benign}: A flow generated with the intention of being benign and/or in relation to a benign activity. E.g., benign user traffic, operating system updates, multicast traffic from a remote benign device, etc.
    \item \textbf{Malicious}: A flow generated with the intention of being harmful, disruptive, and purposefully attacking and/or in relation to an attacking activity. E.g., a port scan to find computers to attack, a denial of service attack, an exploit attempt, or flows that are responses of a non-infected IP address to an attack attempt.
    \item \textbf{Unknown}: A flow whose intention is unknown, or it is hard to decide if it is benign or not.
\end{itemize}

The second and third levels are called \textit{source} and \textit{destination} respectively, and they label the source and destination IP address of the flow. The items for \textit{source} are \textit{From\_malicious}, and \textit{From\_benign}; and for the \textit{destination} are \textit{To\_malicious}, and \textit{To\_benign}.

As a motivating example, imagine a flow from host A to host B, both in the same local network. If the flow is labeled \textit{Malicious}, there is no way to know if A is the malicious host or if B is the malicious host. So the second and third levels of the proposed ontology define labels for the source and destination hosts.

The fourth level of the ontology is the \textit{technique}. This is meant to provide a high-level category for the attack or benign technique used in the flow. The technique is independent of the main label and should include techniques for any of the first-level items. Example items are \textit{DoS}, \textit{Command\_and\_control}, \textit{Discovery}, or \textit{Initial\_access}.

The fifth level of the ontology is the \textit{sub-technique}. This is meant to provide a detailed explanation of the technique used whenever this information is available. The \textit{sub-technique} level is optional, however, if a \textit{technique} level is used, it is recommended to identify the sub-technique. For a fourth level label such as \textit{DoS}, sub-techniques may include distributed DoS (DDoS) or ransom DoS (RDoS), which provide more insight into which type of DoS was observed.

The sixth level of the ontology is the \textit{process} that generated the label. This is meant to provide specific names for the process that generated the flow, which can be a specific malware, operating system, or tool. Example items in this level are \textit{Windows, Linux, or Mirai}.

The seventh level of the ontology is named \textit{app-protocol} for the application level protocol. Even though some network flow generators like Zeek~\cite{noauthor_zeek_nodate} generate columns with the application-level protocol, this is not common for most network monitors. It can also be the case that Zeek misidentifies a protocol, or that the ontology includes protocols that are not common. In the case of misidentification of protocols by the flow creation tool, this level allows the assignment of the correct application-level protocol. If the \textit{app-protocol} level is missing it can be considered that the \textit{app-protocol} is unknown.

All the levels of the ontology can be extended and annotated in its implementation, as shown in Section~\autoref{sec-tool}.

\subsection{Labeling the port scan with the ontology}

Applying the ontology to the port scan labeling use case, each port scan-related flow should be labeled as follows:

\begin{enumerate}
    \item \textbf{label:} Malicious
    \item \textbf{source:} From\_malicious
    \item \textbf{destination:} To\_benign
    \item \textbf{technique:} Discovery
    \item \textbf{sub-technique:} Port\_discovery
    \item \textbf{app-process:} Nmap
\end{enumerate}

Notice that there is no \textit{app-protocol} level because the port scan used TCP packets, and that protocol is already part of the flow. Since the port was not open, then there is no application-level protocol used.

Incidentally, the flow itself may (and should) present more information about the flows of the port scan. In the case of Zeek, it would include the \textit{state} of the flow to recognize if it was established (the port was open) or if it was an attempt (ports were closed). If the port was open and data was exchanged, it may be possible to even have the \textit{service} that can be, for example, SSH.

\section{A tool for labeling network security flows}
\label{sec-tool}

To show the utility of the ontology and how to apply it in a real-life scenario, we developed a tool called \textbf{NetflowLabeler} (software available in \url{https://anonymous.4open.science/r/netflowlabeler-B5F9/}). The labeling rule format is based on the Argus tool called \textit{ralabel} that uses a configuration file and rules to label Argus flows\footnote{https://github.com/openargus/clients/tree/master/examples/ralabel}.

NetflowLabeler is a Python tool that, given a configuration file with rules, it can label Zeek network flow files. The rules for labeling and using the ontology are included in the configuration. The rules are used to first label the Zeek~\textit{conn.log} file and then the rest of the Zeek files. The ontology is also defined in the configuration file so the user can see the options available and expand the ontology if needed (for example, by adding new protocols).  

The rules to apply the labels in the configuration follow a hierarchical and nested order. Rules are of the form:
\begin{verbatim}
label, detailed-label:
    - 1st condition & 2nd condition
    - 3rd condition
    - 4th condition
\end{verbatim}

The logical conjunction AND is created by the keyword \textit{\&} in the same condition line. The logical disjunction OR is created by adding a new line with a new condition. So all condition lines implicitly use \textit{OR}.

Each condition can be created by accessing all the columns of the flows and matching them against numerical conditions with the operators $<$, $>$, $<=$, $>=$, and $=$. The columns to choose from are: Date, start, Duration, Proto, srcIP, srcPort, dstIP, dstPort, State, Tos, Packets, and Bytes.

The main label is appended as a new column to the original columns in a new file. The detailed label is a unique string concatenating all the rest of the labels together using the separator "-" (dash). By default, if any of the two labels are not assigned by the rules, the tool assigns the special text "\textit{(empty)}" since this is the standard procedure for empty fields in Zeek log files.

For example, to label all the flows belonging to a DoS attack that is sent from IP 77.67.96.222 AND using protocol UDP, OR from IP 122.17.49.142 AND using protocol TCP, OR to destination IP 2a00:1450:400c:c05::69, the configuration file would be:

\begin{verbatim}
Malicious, From_malicious-To_benign-
   DoS-DDoS-Linux-NTP:
    - srcIP=77.67.96.222 and Proto=UDP
    - srcIP=122.17.49.142 and Proto=TCP
    - dstIP=2a00:1450:400c:c05::69 
\end{verbatim}

Once the \textit{conn.log} file is labeled, the module \textit{zeek-files-labeler.py} can be used to transfer these labels to the rest of the Zeek log files, such as \textit{ssl.log}, \textit{http.log}, etc.

The transfer is performed using Zeek's \textbf{UID} field, which is shared between interpretations of the same flow in all the files. There are two exceptions. First, the \textit{files.log} file which only has correspondences with protocols transferring files, and the \textit{x509.log} file which only has correspondences with the \textit{ssl.log} file. All the Zeek files are labeled flow by flow.

This rules configuration file should not only greatly improve the assignment of labels, but it can prove crucial for model creators to understand, evaluate and explain their machine learning models.

\subsection{Labeling the port scan with the tool}

The example port scan would be labeled using the following configuration file:

\begin{verbatim}
Malicious, From_malicious-To_benign-
   Discovery-Port_discovery-Linux:
    - srcIP=44.61.93.2 and dstIP
       =192.168.1.100 and Proto=TCP
\end{verbatim}

The configuration file can be more precise, for example, including the specific port ranges that were scanned.

\section{Perspectives and evaluations of defenders}
\label{sec-using-labels}

An often unexplored perspective for evaluating datasets is their usefulness for comparing detection models in real environments. This perspective cannot be addressed only with better labels, as it needs models to be available, datasets to be homogenized, metrics to be agreed upon, and output results to be comparable. This section serves as a thought experiment to consider how far away datasets are as a means for a complete evaluation of tools.

The defender's perspective as an evaluator is to know which detection method is better. This means that under equal conditions and data, the evaluator would like to know which method performs better in the sense of low False Positive Rate (FPR), F1-measure, etc. (the specific metrics are not relevant here). However, in a real scenario, five more evaluations should be considered.

\subsection{Flow or IP address} Most detection methods detect flows or packets as malicious but do not decide if the source IP address or destination IP address is detected. Since most protection mechanisms work by blocking IP addresses or domains, there should be a transformation from individual flow-level detections to IP addresses. This may be solved with a threshold on the ratio of benign vs malicious flows, but it requires estimating the threshold, computing errors, etc. 

The problem is aggravated when multiple alerts based on flows from the same attack should be transformed into one unique high-level alert for the defender to work on. This is related to the alert fatigue problem~\cite{radebe_perceptions_2022}, and it may need more machine learning algorithms to solve. 

\subsection{Detection and undetection} The second perspective is about \textit{when} an IP address is detected, and \textit{when} it should be undetected. Imagine that an IP address is attacking from 8 am to 9 am, then \textit{not} attacking from 9 am to 10 am, and attacking \textit{again} from 10 am. The defender would need one detection before 9 am to block the IP address, after 9 am it should be shown that the IP stopped attacking and may be {undetected}, and after 10 am it should be detected again. Labels in the dataset should be able to allow such evaluation by having labels with enough precision and timing. 

\begin{figure}[h]
\caption{Example of detection of malicious Host A in time. Comparison of the flows involved in the detection. Even though at the IP level it was correctly detected, at the flow level, some malicious flows were missed, and some benign flows were misused.}
\centering
\includegraphics[width=0.5\textwidth]{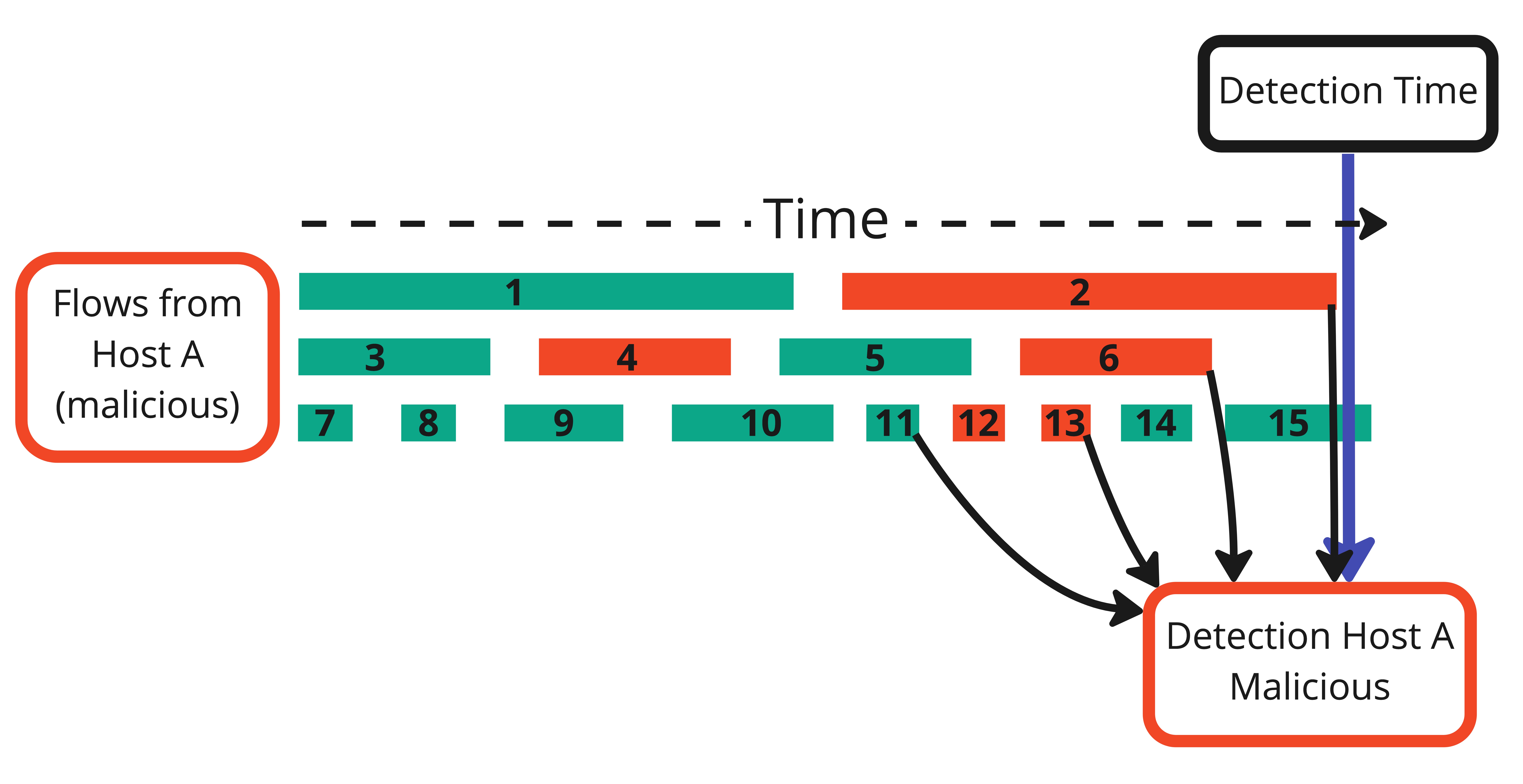}
\label{fig-flowstimeline}
\end{figure}

On top of this second perspective is the evaluation of metrics (e.g., FPR) per IP address vs. per flow. A method with good flow detection metrics is not necessarily good with IP address detection metrics. This is shown better in~\autoref{fig-flowstimeline}, where malicious Host A can be seen sending many flows, both benign (green) and malicious (red). The detection of the malicious host happens at some moment \textit{Detection Time}, and it is based on four flows referenced with black arrows. Host A is correctly detected, so it is a true positive at the IP address level. From the 15 flows sent by Host A up to that point (numbers 1 to 15), there were 5 malicious (in red numbers 2, 4, 6, 12, and 13) and 10 benign (in green). Of the 5 malicious flows, only 3 were correctly used in the detection, numbers 2, 6, and 13. Of the 10 benign flows, number 11 was mistakenly used as evidence for the detection of Host A.

Computing the performance of ~\autoref{fig-flowstimeline} gives a $FPR=10\%$, a $F1-score=66\%$, an $Accuracy=80\%$, and a $TPR=60\%$ at the flow level. This result suggests that a poor-performing detector at the flow level can still be a good detector at the IP address level.

\subsection{Blocking} The third perspective is about the transition from IP address detections to IP blocking. Blocking is a critical action when traffic is stopped to or from the alerted IP address. This decision is important and prone to errors and few real-life systems implement it automatically, leaving the decision to humans. The main reason is that a few false positives can have strong consequences for an organization.

\subsection{When detection happens} The timing of the detection is the fourth perspective. The defender needs methods that detect the threats as fast as possible. Therefore \textit{when} the method detects a threat is crucial. At first, it may seem that this is possible to evaluate since datasets have time values for flows and packets. However, an IP address may be detected because it generated 30 alerts in the last hour. A method that can alert about an IP address after 10 detections would be better than a method that needs 20 detections. However, this should always be balanced with the detection performance (e.g., FPR) since a good detection performance is better than an \textit{early} detection. 

\subsection{The correct detection reasons} The fifth perspective is about detecting the correct attack and not just \textit{any} attack. This is commonly referred to as Type IV error or arriving at the correct result for the incorrect reasons~\cite{ottenbacher_statistical_1992}. Defenders working to detect ransomware, for example, need to know if an IP address was detected and blocked due to ransomware alerts and not because of a DDoS attack. This is a crucial evaluation, and it can only be done if the labels of the dataset are detailed enough, which is not in most datasets.

\subsection{The defender's perspective on a port scan}
Using the case of the port scan, the defender would like to detect it without errors. However, transitioning from flow labels to blocking the attacker is not straightforward. We consider here each of the five previously described perspectives for a defender.

In the first perspective, from flows to IP address detections, the defender wants to know if the \textit{IP address} is malicious because some flows have been detected as port scans. The correct detection requires detailed labels. It is also related to the confidence of the detector and to other contextual information. 

In the second perspective, when an IP address is detected, we want to alert and block it as fast as possible, but we do not want an alert per \textit{port scan} flow since there may be thousands. So there should be a balance between having enough information to detect it with confidence, not overloading the defender with alerts, and not forgetting the number of alerts. This last part is important to distinguish a short port scan from a massive port scan.

In the third perspective, blocking a port scan, there is more confidence in blocking an incoming port scan but not so much in blocking an outgoing port scan since many behaviors can be detected as port scans. This suggests that the control of false positives can also be done with contextual information on the local network or IP address, which is not present in most datasets. However, our label ontology can help if the port scans from administrators are labeled "Benign, From\_benign-To\_benign-Discovery-Port\_discovery".

In the fourth perspective, the defender wants the detection as fast as possible. However, since detection methods have false positives it is usually desired a more confident detection, with more than one port scan flow detected, before the IP address is determined to be malicious. Otherwise, many false positive alerts will be recorded. On the other hand, if a detection method waits too much and the port scan is alerted very late, it would be a very confident but probably useless detection.

In the fifth perspective, detecting the IP address for the correct reasons, it is important to know if a port scan was enough to alert and block the IP address (maybe because it was a massive port scan) or if it was not enough and the IP address should not be blocked. But it must be clear that the IP address was not alerted and blocked as a port scanner because of another type of attack.

All the above perspectives need to be taken into account from the defenders' point of view. However, binary-labeled datasets do not provide the necessary information for the evaluation of the detectors. More detailed labels can help alleviate some of the problems even though they cannot address all of the issues.

\section{Conclusion}
\label{sec-conclusion-labels}
The design and assignment of labels in network security datasets are difficult topics both for dataset creators and dataset users. A lot of effort has been done to improve the datasets and the models, but still, a lot of work is needed to improve the assignment of labels. Our analysis of the needs for labels from different stakeholders guided the design of a new dataset labeling ontology that may help dataset creators to correctly assign labels with different levels of specificity. The ontology is implemented in a free-software tool that reads a label configuration file to label all Zeek log files. We argue that the ontology-based rules configuration file used to label a dataset should be published as part of the artifacts of a dataset in order for others to understand, reproduce and evaluate the dataset.



\section*{Data and Software Availability}
\label{sec-data-software-availability}
The complete ontology and tool have been published in \url{https://anonymous.4open.science/r/netflowlabeler-B5F9/}


\bibliographystyle{plain}
\bibliography{references}

\appendices

\end{document}